\documentclass{ws-fnl}
\usepackage{cite}
\begin{document}

\markboth{D.\ S.\ Goldobin}{Distribution of Watanabe--Strogatz variables \& Circular Cumulants}

\catchline{}{}{}{}{}

\title{
Relationships between the Distribution of Watanabe--Strogatz Variables and Circular Cumulants for Ensembles of Phase Elements}

\author{Denis S.\ Goldobin}

\address{Institute of Continuous Media Mechanics, UB RAS, Academician Korolev Street 1\\
614013 Perm, Russia,\\
Department of Theoretical Physics, Perm State University, Bukirev Street 15\\
614990 Perm, Russia\\
Denis.Goldobin@gmail.com}

\maketitle

\begin{history}
\received{(received date)}
\revised{(revised date)}
\end{history}

\begin{abstract}
The Watanabe--Strogatz and Ott--Antonsen theories provided a seminal framework for rigorous and comprehensive studies of collective phenomena in a broad class of paradigmatic models for ensembles of coupled oscillators.
Recently, a ``circular cumulant'' approach was suggested for constructing the perturbation theory for the Ott--Antonsen approach.
In this paper, we derive the relations between the distribution of Watanabe--Strogatz phases and the circular cumulants of the original phases.
These relations are important for the interpretation of the circular cumulant approach in the context of the Watanabe--Strogatz and Ott--Antonsen theories.
Special attention is paid to the case of hierarchy of circular cumulants, which is generally relevant for constructing perturbation theories for the Watanabe--Strogatz and Ott--Antonsen approaches.

\vspace{10pt}\noindent{\it Keywords:}
Watanabe-Strogatz theory;
Ott-Antonsen theory;
circular cumulants.
\end{abstract}

\section{Introduction}
The interest to dynamics of coupled oscillators is related to many applications in physics, biology and engineering~\cite{Acebron-etal-2005,Pikovsky-Rosenblum-2015}. In the case of weak coupling, one can develop an universal approach based on the phase reduction, where only the dynamics of phases is considered while the amplitudes are functions of phases. Famous Kuramoto model describes the system of phase oscillators coupled via the mean field; this system allows for an analytical description of the synchronization transition. For certain class of phase systems in common field (see the next section for specific definition and review~\cite{Pikovsky-Rosenblum-2015}), like the Kuramoto model, Watanabe and Strogatz (WS)~\cite{Watanabe-Strogatz-1993,Watanabe-Strogatz-1994,Marvel-Mirollo-Strogatz-2009} and Ott and Antonsen (OA)~\cite{Ott-Antonsen-2008} developed analytical approaches.

Within the WS approach, for ensembles of identical elements, phases $\varphi_k$ can be mapped into auxiliary phases $\psi_k$ [via M\"obius transformation, see Eq.~(\ref{eq104})]; the distribution of $\psi_k$ is frozen and only the complex-valued mapping parameter can evolve non-trivially in time. Synchronization transition can be characterized with this mapping parameter, depending on which the frozen set of $\psi_k$ maps into a squeezed or spread set of original phases $\varphi_k$. The OA theory represents a different approach, which yields the evolution equations for the complex-valued order parameter $N^{-1}\sum_{k=1}^Ne^{i\varphi_k}$ in the thermodynamic limit $N\to\infty$.

Recently~\cite{Tyulkina-etal-2018,Goldobin-etal-2018,Tyulkina-etal-2018b}, the so-called ``circular cumulant'' approach was suggested as a framework for constructing the perturbation theories for systems where the conditions of the WS and OA theories are imperfectly satisfied. This recent approach significantly differs for the perturbation theory developed in~\cite{Vlasov-etal-2016} on the basis of the WS variables; in particular, it does not encounter singularity issues for high-synchrony regimes. Both comparison of two approaches and interpretation of the circular cumulant representation necessitate the task of this paper, which is to interpret the circular cumulants in terms of the WS theory. Early, interpretation and deeper understanding of the WS and OA theories 
received significant attention in the literature~\cite{Pikovsky-Rosenblum-2008,Marvel-Mirollo-Strogatz-2009,Pazo-Montbrio-2014,Montbrio-Pazo-Roxin-2015,Chen-Engelbrecht-Mirollo-2017}.

The paper is organized as follows. In Sec.~\ref{sec:prelim}, we recall the basic information from the WS and OA theories, which is required for interpretation of circular cumulants in terms of the WS theory. In Sec.~\ref{sec:res}, the relations between circular cumulants and the WS variables are reported. In Sec.~\ref{sec:concl}, the conclusions are drawn. In Appendix, we provide the details of derivations.

\section{Mathematical preliminaries}
\label{sec:prelim}
\subsection{Watanabe--Strogatz and Ott--Antonsen approaches}
The dynamics of the ensemble of $N$ identical phase oscillators of the from
\begin{equation}
\dot\varphi_k=\Omega(t)+\mathrm{Im}(2h(t)e^{-i\varphi_k})\,,
\label{eq101}
\end{equation}
where $k=1,...,N$, is known to possess $N-3$ integrals of motion~\cite{Watanabe-Strogatz-1993, Watanabe-Strogatz-1994, Pikovsky-Rosenblum-2008, Marvel-Mirollo-Strogatz-2009
}
and to be governed by one ODE for a complex variable $z$:
\begin{equation}
\dot{z}=i\omega(t)\,z+h(t)-h^\ast(t)\,z^2.
\label{eq102}
\end{equation}
The growth rates of auxiliary phases $\psi_k$, known as Watanabe--Strogatz (WS) variables, are identical for all phase elements:
\begin{equation}
\dot{\psi}_k=\omega(t)+\mathrm{Im}(2h(t)\,z^\ast)\,.
\label{eq103}
\end{equation}
The variables $z$ and $\{\psi_k\}$ are uniquely defined by the relations
\begin{equation}
e^{i\varphi_k}=\frac{z+e^{i\psi_k}}{1+z^\ast e^{i\psi_k}}
\qquad\mbox{or, inverted relation, }\qquad
e^{i\psi_k}=\frac{e^{i\varphi_k}-z}{1-z^\ast e^{i\varphi_k}}
\label{eq104}
\end{equation}
under the condition $\sum_{k=1}^Ne^{i\psi_k}=0$\,.

In the thermodynamic limit $N\to\infty$, it is natural to consider the ensemble dynamics in terms of the probability density $w(\varphi,t)$. The dynamics of $w(\varphi,t)$ is governed by the Master-equation
\begin{equation}
\frac{\partial w}{\partial t}
 +\frac{\partial}{\partial\varphi}\Big((\Omega-ihe^{-i\varphi}+ih^\ast e^{i\varphi})w\Big)
 =0\,.
\label{eq105}
\end{equation}
In the Fourier space, where $w(\varphi,t)=(2\pi)^{-1}[1+\sum_{j=1}^{\infty}(a_je^{-ij\varphi}+c.c.)]$, the Master-equation reads
\begin{equation}
\dot{a}_{j}=ji\Omega a_j+jh{a}_{j-1}-jh^\ast{a}_{j+1},
\label{eq106}
\end{equation}
for $j\ge1$, where $a_0=1$.

In~\cite{Ott-Antonsen-2008}, Eq.~(\ref{eq106}) was reported to admit particular solutions of the form $a_j=(a_1)^j$ with $a_1$ obeying the equation
\begin{equation}
\dot{a_1}=i\Omega a_1+h-h^\ast a_1^2\,.
\label{eq107}
\end{equation}
The manifold of $a_j=(a_1)^j$ is referred to as Ott--Antonsen (OA) manifold.
The usefulness of this result can be seen from the fact that, in the thermodynamic limit, $a_1$ is the order parameter, $Re^{i\Phi}\equiv\langle{e^{i\varphi}}\rangle=a_1$. The latter provides opportunity for a comprehensive and rigorous study of diverse collective phenomena in ensembles of phase elements (e.g., see~\cite{Pikovsky-Rosenblum-2008, Pazo-Montbrio-2014, Montbrio-Pazo-Roxin-2015, Abrams-etal-2008, Nagai-Kori-2010, Braun-etal-2012, Pimenova-etal-2016, Goldobin-etal-2017, Dolmatova-etal-2017, Zaks-Tomov-2016}).

In terms of Watanabe--Strogatz variables for $N\to\infty$, variable $z$ is still governed by Eq.~(\ref{eq102}) and the corresponding probability density $W(\psi,t)$ is a frozen wave of an arbitrary shape propagating with velocity~(\ref{eq103}). Even though Eqs.~(\ref{eq107}) and (\ref{eq102}) are similar, generally $z$ is not an order parameter and the calculation of the order parameter $a_1$ from $z$ and $W(\psi)$ is a laborious task. The OA solution in terms of the WS variables corresponds to the case of $W(\psi,t)=(2\pi)^{-1}$, $z=a_1$. Notice, that the OA manifold is neutrally stable for perfectly identical elements, since $W(\psi,t)$ is a frozen wave but not attracted to the uniform state. However, the OA approach can be generalized to certain cases of ensembles with nonidentical parameters (frequencies $\Omega(t)$, or $h(t)$); in situations of practical interest, the nonidentities make the OA manifold attracting~\cite{Ott-Antonsen-2008,Pietras-Daffertshofer-2016}. Since in reality the identity of elements is never perfect, the OA solutions are attracting.

\subsection{Circular cumulant approach}
The description of the ensemble dynamics in the vicinity of the OA manifold in terms of $a_j$ is problematic for high-synchrony states, where $|a_1|$ is close to $1$ and the series $a_j\sim(a_1)^j$ possesses a poor convergence properties. In this case it can be more efficient to go from considering moments $a_j=\langle e^{ij\varphi_k}\rangle$ to the formally-cor\-respond\-ing cumulants $K_j$ determined by the generating functions:
\begin{equation}
F(\zeta)\equiv\langle\exp(\zeta e^{i\varphi_k})\rangle \equiv\sum_{j=0}^{\infty}a_j\frac{\zeta^j}{j!}\,,
\qquad
\ln(F(\zeta))\equiv\sum_{j=1}^{\infty}K_j\frac{\zeta^j}{j!}\,.
\label{eq201}
\end{equation}

In terms of $K_j$, Eqs.~(\ref{eq106}) take the form
\[
\dot{K}_j=ji\Omega K_j+h\delta_{1j}
 -jh^\ast\Big(K_{j+1}+\sum_{m=1}^j\frac{(j-1)!}{(m-1)!\,(m-j)!}K_{j-m+1}K_{m}\Big)\,,
\]
where $\delta_{1j}=1$ for $j=1$ and zero otherwise. The derivation of equations for $K_n$ from Eqs.~(\ref{eq106}) can be found in~\cite{Tyulkina-etal-2018}; similar derivation can be performed also for some other cases, deviating from the form~(\ref{eq106}) [or Eq.~(\ref{eq101})].

For specific physical systems, it is frequently more convenient to use
\[
\varkappa_j\equiv\frac{K_j}{(j-1)!}
\]
(see~\cite{Tyulkina-etal-2018,Goldobin-etal-2018}); the governing equations for $\varkappa_j$ read
\begin{equation}
\dot{\varkappa}_n=ni\Omega\varkappa_n+h\delta_{1n}
 -nh^\ast\big(n\varkappa_{n+1}
 +{\textstyle\sum_{m=1}^n}\varkappa_{n-m+1}\varkappa_m\big)\,.
\label{eq202}
\end{equation}
We refer to $\varkappa_j$ as to ``circular cumulants''.

One of the most generic and important violations of the OA form~(\ref{eq101}) is the case of intrinsic (individual) noise acting on the phase elements. With this noise, Eq.~(\ref{eq101}) acquires the form
\begin{equation}
\dot\varphi_k=\Omega(t)+\mathrm{Im}(2h(t)e^{-i\varphi_k})+\sigma\xi_k(t)\,,
\label{eq-s1}
\end{equation}
where $\sigma$ is the strength of intrinsic noise, $\xi_k$ are independent normalized $\delta$-correlated Gaussian noise signals: $\langle\xi_k\rangle=0$, $\langle\xi_k(t)\,\xi_m(t')\rangle=2\delta_{km}\delta(t-t')$\,. Eq.~(\ref{eq106}) changes to
\begin{equation}
\dot{a}_{j}=ji\Omega a_j+jh{a}_{j-1}-jh^\ast{a}_{j+1}-j^2\sigma^2a_j,
\label{eq-s2}
\end{equation}
which does not admit the OA ansatz $a_j=(a_1)^j$, but can be treated within the framework of the cumulant approach~\cite{Tyulkina-etal-2018} with Eq.~(\ref{eq202}) modified to
\begin{align}
\dot{\varkappa}_n=ni\Omega\varkappa_n+h\delta_{1n}
 -nh^\ast\big(n\varkappa_{n+1}
 +{\textstyle\sum_{m=1}^n}\varkappa_{n-m+1}\varkappa_m\big)\qquad
\nonumber\\
-\sigma^2n\big(n\varkappa_{n}
 +{\textstyle\sum_{m=1}^{n-1}}\varkappa_{n-m}\varkappa_m\big)
 \,.
\label{eq-s3}
\end{align}

\begin{figure}
\centering
\includegraphics[width=0.5\columnwidth]%
 {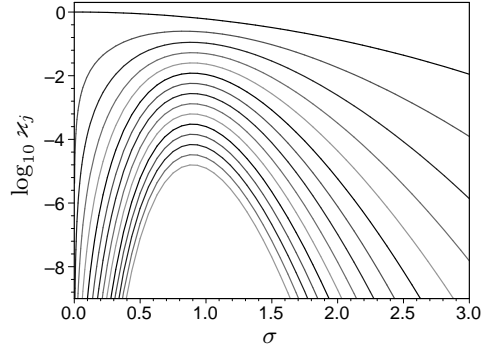}
\caption{First 15 circular cumulants $\varkappa_j$ for the wrapped Gaussian distribution of width $\sigma$.}
\label{fig1}
\end{figure}

For constructing the perturbation theories on top of the Ott--Antonsen theory, there are several benefits:
\\
$\bullet$~For the OA solution, $\varkappa_1=a_1$ and all higher circular cumulants $\varkappa_{j\ge2}=0$.
\\
$\bullet$~While $a_j$ converges poorly for $|a_1|\to1$, with circular cumulants, $|\varkappa_1|\to1$ requires $|\varkappa_{j\ge2}|\to0$ and the issue of convergence for series $\varkappa_j$ does not arise.
\\
$\bullet$~Intrinsic noise of strength $\sigma$ 
creates for the circular cumulants hierarchy of smallness $\varkappa_j\propto\sigma^{2(j-1)}$~\cite{Tyulkina-etal-2018,Goldobin-etal-2018}. Moreover, the wrapped Gaussian distribution for phases with $a_1=e^{i\Phi-\sigma^2/2}$ and $a_j=e^{ij\Phi-\sigma^2j^2/2}$, which emerges in some cases where the OA form~(\ref{eq101}) is violated~\cite{Hannay-etal-2018}, generates a well pronounced hierarchy for arbitrary value of $\sigma$ (see Fig.~\ref{fig1}). Thus, the formation of hierarchies of $\varkappa_j$ is frequent in specific physical problems; the presence of such a hierarchy is a favorable condition for constructing analytical approximations.

Notice, in Fig.~\ref{fig1} for $\sigma^2\ll1$, one can see the hierarchy $\varkappa_j\propto\sigma^{2(j-1)}$; for $\sigma^2\gg1$, $\varkappa_j\propto(e^{-\sigma^{2}/2})^j$; for moderate values of $\sigma^2$, where there is no small parameter, some intermediate hierarchy with law $|\varkappa_{j+1}/\varkappa_j|\approx\varepsilon$ can be still observed numerically (here $\varepsilon$ can attain the maximal value of ca.\ $0.462$). For the Kuramoto ensemble with Gaussian intrinsic noise and Lorentzian distribution of individual oscillator frequencies, similar picture is observed~\cite{Goldobin-etal-2018} for steady state regimes; where one can identify some small parameter $\varepsilon$, the hierarchies of sort $\varkappa_j\propto\varepsilon^{j-1}$ or $\varkappa_j\propto\varepsilon^j$ form, and, with no small parameter, some intermediate but well pronounced geometric progressions are observed for $\varkappa_j$. In Ref.~\cite{Tyulkina-etal-2018}, where the violation of the general OA form~(\ref{eq101}) was caused by the Gaussian intrinsic noise of intensity $\sigma^2$, the hierarchy $\varkappa_j\propto\sigma^{2(j-1)}$ was reported for perturbed OA solutions. On these grounds, the latter case of hierarchy will be of special interest for us, while we will keep in mind that other sorts of hierarchies are also possible.

We cannot prove it rigorously, but formulate it as an important observation: the law $|\varkappa_{j+1}/\varkappa_j|\approx\varepsilon<1$, generalizing all these hierarchies, is very frequent. The presence of a hierarchy is not as constraining as specific closures for low-dimensional reductions (like Ott--Antonsen ansatz or Gaussian approximation): with a hierarchy the cumulants remain uncertain up to a factor of the order of magnitude of 1.

\section{Results}
\label{sec:res}
\subsection{Calculation of the density of Watanabe--Strogatz variables from circular cumulants} 
Firstly, we establish relations between $W(\psi)$ and $w(\varphi)$, representing the former in terms of $A_j=\int_0^{2\pi}W(\psi)\,e^{ij\psi}\mathrm{d}\psi$ and the latter in terms of circular cumulants $\varkappa_j$ or $K_j=(j-1)!\,\varkappa_j$. In Fourier space, $W(\psi)=(2\pi)^{-1}[1+\sum_{j=1}^\infty(A_je^{-ij\psi}+c.c.)]$, where `$c.c.$' stands for complex conjugate.

Eq.~(\ref{eq104}) yields
\begin{equation}
\int\mathrm{d}\psi\,W(\psi)\,e^{im\psi}
=\int\mathrm{d}\varphi\,w(\varphi)\left(\frac{e^{i\varphi}-z}{1-z^\ast e^{i\varphi}}\right)^m\,,
\label{eqr00}
\end{equation}
and one can find (see Appendix~\ref{sec:der-Aj} for the derivation)
\begin{equation}
A_j=\big(1+\sum_{m=2}^\infty p_m\widehat{Q}_m\big)A_j^{(0)},
\label{eqr01}
\end{equation}
where
\[
\widehat{Q}_m\equiv\frac{1}{m!}\left(\frac{\partial}{\partial K_1}\right)^m;
\]
$p_j$ are moments corresponding to cumulants $K_j$ with removed $K_1$---specifically,
\begin{align}
p_1&=0\,,
\nonumber\\
p_2&=K_2\,,
\nonumber\\
p_3&=K_3\,,
\nonumber\\
p_4&=K_4+3K_2^2\,,
\nonumber\\
p_5&=K_5+10K_3K_2\,,
\nonumber\\
p_6&=K_6+15K_4K_2 +10K_3^2 +15K_2^3\,,
\nonumber\\
&\dots\,,
\label{eqr02}
\end{align}
in other words, $p_j$ correspond to moments $a_j$ subject to the transformation of the removal of the first moment $a_1$; and
\begin{equation}
A_{j}^{(0)}=\left(\frac{K_1-z}{1-z^{\ast}K_1}\right)^j.
\label{eqr03}
\end{equation}

Along with $A_j$, one must calculate $z$.
For $W(\psi)$, the condition $A_1=0$ dictates the value of $z$; this condition can be written as
\begin{align}
z-K_1=\sum_{j=2}^\infty p_j\frac{(z^{\ast})^{j-1}(1-|z|^2)}{(1-z^\ast K_1)^{j}}\,.
\label{eqr04}
\end{align}
From condition~(\ref{eqr04}), one can iteratively calculate $z$ with any required accuracy. With diverse hierarchies for $\varkappa_j$, there will be fast decaying hierarchies of $p_j$ providing a fast convergence for scheme
\begin{align}
z_0&=K_1\,,
\qquad
z_1
=K_1+\frac{K_1^\ast K_2}{1-|K_1|^2}\,,
\qquad
\dots\,,
\nonumber\\
z_n&=K_1+\sum_{j=2}^{n+1}p_j\frac{(z_{n-1}^{\ast})^{j-1}(1-|z_{n-1}|^2)}
  {(1-z_{n-1}^\ast K_1)^{j}}\,,
\qquad
\dots\,.
\label{eqr05}
\end{align}
The exact value $z=z_n+\mathcal{O}(p_{n+2})$ or, as one can see from Eq.~(\ref{eqr02}),
\[
z=z_n+\left\{
\begin{array}{cl}
\mathcal{O}(\varepsilon^{\mathrm{ceil}(n/2)+1})\,,
&\mbox{ for }\varkappa_j\propto\varepsilon^{j-1}\,;
\\
\mathcal{O}(\varepsilon^{n+2})\,,
&\mbox{ for }\varkappa_j\propto\varepsilon^j\,.
\end{array}
\right.
\]
Here function $\mathrm{ceil}(x)$ is the smallest integer greater than or equal to $x$.

Specifically for the hierarchy $\varkappa_j\propto\varepsilon^{j-1}$, one can calculate: $z\approx Z_0=z^{(0)}=K_1$ (with $\varepsilon$-contributions neglected); $z\approx Z_1=z^{(0)}+z^{(1)}$ (with $\varepsilon^2$-contributions neglected), where
\[
z^{(1)}=\frac{K_1^\ast K_2}{1-|K_1|^2}\,;
\]
$z\approx Z_2=z^{(0)}+z^{(1)}+z^{(2)}$ (with $\varepsilon^3$-contributions neglected), where
\[
z^{(2)}=\frac{K_3K_1^{\ast 2}}{(1-|K_1|^2)^2}
 +\frac{2K_1^{\ast 3}K_2^2+K_1|K_2|^2}{(1-|K_1|^2)^3}\,;
\]
$z\approx Z_3=z^{(0)}+z^{(1)}+z^{(2)}+z^{(3)}$ (with $\varepsilon^4$-contributions neglected), where
\begin{align}
z^{(3)}&=\frac{K_4K_1^{\ast 3}}{(1-|K_1|^2)^3}
 +\frac{K_3(8K_2K_1^{\ast 4}+2K_2^{\ast}|K_1|^2)+K_3^{\ast}K_2K_1^2}{(1-|K_1|^2)^4}
\nonumber\\
&\qquad
 +\frac{10K_2^3K_1^{\ast 5}+(1+6|K_1|^2)K_2|K_2|^2K_1^{\ast}+3K_2^\ast|K_2|^2K_1^3}{(1-|K_1|^2)^5}\,.
\nonumber
\end{align}
Here calculation of $Z_n$ corresponds to calculation of $z_{2n-1}$ (\ref{eqr05}) for $n\ge1$.

With Eq.~(\ref{eqr03}), one can find a recurrence relation:
\begin{align}
\frac{\partial A_j^{(0)}}{\partial K_1}&
 =\frac{\partial}{\partial K_1}\left(\frac{K_1-z}{1-z^{\ast}K_1}\right)^j
 =j\frac{(1-|z|^2)(K_1-z)^{j-1}}{(1-z^{\ast}K_1)^{j+1}}
\nonumber\\
&=\frac{j}{1-|z|^2}\left(z^{\ast 2}A_{j+1}^{(0)}+2z^{\ast}A_{j}^{(0)}+A_{j-1}^{(0)}\right)\,.
\label{eqr06}
\end{align}
Note, $A_0^{(0)}=1$ and $(\partial A_1^{(0)}/\partial K_1)\ne0$ (the condition $A_1=0$ is fulfilled for certain $z$ and $K_1$, which does not mean that partial derivatives of $A_1^{(0)}$ have to be $0$).

\subsection{Scaling laws of order parameters $A_j$ of Watanabe--Strogatz variables for hierarchy of circular cumulants $\varkappa_j\propto\varepsilon^{j-1}$} 
We explicitly indicate the hierarchy with notation $\varkappa_j\equiv\varepsilon^{j-1}s_{j-1}$.
Since $z-K_1\sim\varepsilon$, $A_j^{(0)}\propto\varepsilon^{j}$; employing the recurrence relation~(\ref{eqr06}), to the leading order, one finds
\begin{align}
\widehat{Q}_mA_j^{(0)}&\approx\frac{j!}{m!\,(j-m)!}\frac{A_{j-m}^{(0)}}{(1-|z|^2)^m}
=\frac{C_m^j(s_0-z)^{j-m}}{(1-|z|^2)^m(1-z^{\ast}s_0)^{j-m}}&
\nonumber\\
&\approx C_m^j\frac{(-\varepsilon s_0^{\ast}s_1)^{j-m}}{(1-|s_0|^2)^{2j-m}}\,,&
\mbox{ for }m\le j\,,
\nonumber\\
\widehat{Q}_mA_j^{(0)}&\propto(z^\ast)^{m-j}A_0^{(0)}\propto(z^\ast)^{m-j}\,,&
\mbox{ for }m>j\,
\nonumber
\end{align}
where $C_m^j\equiv{j!/[m!(j-m)!]}$.
With $p_j\propto\varepsilon^{\mathrm{ceil}(j/2)}$, keeping only the leading-order terms in the expression for $A_j$, one can obtain
\begin{align}
A_j&\approx
\left\{
\begin{array}{cr}
\sum_{m=j-1}^{j+1}p_m\widehat{Q}_mA_j^{(0)}, & \mbox{for odd }j\,;\\
p_j\widehat{Q}_jA_j^{(0)}\,, & \mbox{for even }j\,.
\end{array}\right.
\nonumber\\
&\approx
\left\{
\begin{array}{lr}
\frac{\varepsilon^{(j+1)/2}}{(1-|s_0|^2)^j}\Big(
\frac{-js_0^\ast s_1}{1-|s_0|^2}\frac{p_{j-1}}{\varepsilon^{(j-1)/2}}
+\frac{p_j}{\varepsilon^{(j+1)/2}}
+\frac{js_0^\ast}{1-|s_0|^2}\frac{p_{j+1}}{\varepsilon^{(j+1)/2}}
\Big), & \mbox{for odd }j\,;\\
\qquad
\frac{\varepsilon^{j/2}}{(1-|s_0|^2)^j}\frac{p_j}{\varepsilon^{j/2}}\,, & \mbox{for even }j\,.
\end{array}\right.
\label{eqr07}
\end{align}
See Sec.~\ref{sec:der-QmAj} for calculation of $\widehat{Q}_{j}A_j^{(0)}$ and $\widehat{Q}_{j+1}A_j^{(0)}$.
For odd $j$,
\[
p_j\approx\frac{(j-1)j!!}{6}K_3K_2^\frac{j-3}{2}=\varepsilon^{(j+1)/2}\frac{(j-1)j!!}{3}s_2s_1^\frac{j-3}{2}
\]
and, for even $j$,
\[
p_j\approx(j-1)!!\,K_2^{j/2}=\varepsilon^{j/2}(j-1)!!\,s_1^{j/2}\,;
\]
therefore,
\begin{equation}
A_j\approx
\left\{
\begin{array}{cl}
\frac{\varepsilon^{(j+1)/2}j!!(j-1)}{3(1-|s_0|^2)^j}\left(s_2 +\frac{3s_0^{\ast}}{1-|s_0|^2}s_1^2\right)s_1^{(j-3)/2},
 &\mbox{ for odd }j\,;\\[7pt]
\frac{\varepsilon^{j/2}(j-1)!!}{(1-|s_0|^2)^j}s_1^{j/2},
 &\mbox{ for even }j\,.
\end{array}
\right.
\label{eqr08}
\end{equation}
Notice, that $A_j\propto\varepsilon^{\mathrm{ceil}(j/2)}$, while $A_j^{(0)}\propto\varepsilon^{j}$.

\subsubsection{Circular cumulants of the density of Watanabe--Strogatz variables}
Cumulants $\mathcal{K}_j$ of $e^{i\psi}$ with $\langle{e^{i\psi}}\rangle=0$ are
\begin{align}
\mathcal{K}_2&=A_2\,,
\nonumber\\
\mathcal{K}_3&=A_3\,,
\nonumber\\
\mathcal{K}_4&=A_4-3A_2^2\,,
\nonumber\\
\mathcal{K}_5&=A_5-10A_3A_2\,,
\nonumber\\
\mathcal{K}_6&=A_6-15A_4A_2-10A_3^2+30A_2^3\,,
\nonumber\\
&\dots\,.
\nonumber
\end{align}
Substituting \eqref{eqr08}, one finds
\begin{align}
\mathcal{K}_2&=\varepsilon\frac{s_1}{(1-|s_0|^2)^2}
 +\mathcal{O}(\varepsilon^2)\,,
\nonumber\\
\mathcal{K}_3&=\varepsilon^2\frac{2}{(1-|s_0|^2)^3}\left(s_2+\frac{3s_1^2s_0^{\ast}}{1-|s_0|^2}\right) +\mathcal{O}(\varepsilon^3)\,,
\nonumber\\
\mathcal{K}_4&=\varepsilon^2\cdot0
 +\varepsilon^3\frac{3!}{(1-|s_0|^2)^4}\bigg(s_3+\frac{4\cdot2s_2s_1s_0^\ast}{1-|s_0|^2}
 +\frac{12s_1^3s_0^{\ast 2}}{(1-|s_0|^2)^2}\bigg)
 +\mathcal{O}(\varepsilon^4)\,,
\nonumber\\
\mathcal{K}_5&=\varepsilon^3\cdot0+\varepsilon^4\frac{4!}{(1-|s_0|^2)^5}\bigg(
 s_4+\frac{5(2s_3s_1+s_2^2)s_0^\ast}{1-|s_0|^2}
\nonumber\\
&\qquad{}
 +\frac{55s_2s_1^2s_0^{\ast 2}}{(1-|s_0|^2)^2}
 +\frac{55s_1^4s_0^{\ast 3}}{(1-|s_0|^2)^3}\bigg)+\mathcal{O}(\varepsilon^5)\,,
\nonumber\\
\mathcal{K}_6&=\varepsilon^3\cdot0+\varepsilon^4\cdot0 +\varepsilon^5\frac{5!}{(1-|s_0|^2)^6}\bigg(s_5 +\frac{6(2s_4s_1+2s_3s_2)s_0^{\ast}}{1-|s_0|^2}
\nonumber\\
&\quad{}
 +\frac{26(3s_3s_1^2+3s_2^2s_1)s_0^{\ast 2}}{(1-|s_0|^2)^2}
 +\frac{91\cdot4s_2s_1^3s_0^{\ast 3}}{(1-|s_0|^2)^3} +\frac{273s_1^5s_0^{\ast 4}}{(1-|s_0|^2)^4}\bigg)+\mathcal{O}(\varepsilon^6)\,,
\nonumber\\
&\dots\,.
\nonumber
\end{align}
Eq.~\eqref{eqr08} is sufficient only for calculation of the leading term in expression for $\mathcal{K}_j$, but the leading terms are $0$ for $j\ge4$. The provided results were computed with Maple Software and usage of $z=z_n$ calculated with recurrence~(\ref{eqr05}).

One can introduce $\mathcal{S}_j$: $\mathcal{K}_j=\varepsilon^{j-1}(j-1)!\,\mathcal{S}_{j-1}$;
\begin{align}
\mathcal{S}_1&=\frac{s_1}{(1-|s_0|^2)^2}
 +\mathcal{O}(\varepsilon)\,,
\nonumber\\
\mathcal{S}_2&=\frac{1}{(1-|s_0|^2)^3}\left(s_2 +\frac{3s_1^2s_0^{\ast}}{1-|s_0|^2}\right)+\mathcal{O}(\varepsilon)\,,
\nonumber\\
\mathcal{S}_3&=\frac{1}{(1-|s_0|^2)^4}\bigg(s_3+\frac{4\cdot2s_2s_1s_0^\ast}{1-|s_0|^2}
 +\frac{12s_1^3s_0^{\ast 2}}{(1-|s_0|^2)^2}\bigg)
 +\mathcal{O}(\varepsilon)\,,
\nonumber\\
\mathcal{S}_4&=\frac{1}{(1-|s_0|^2)^5}\bigg(
 s_4+\frac{5(2s_3s_1+s_2^2)s_0^\ast}{1-|s_0|^2}
 +\frac{55s_2s_1^2s_0^{\ast 2}}{(1-|s_0|^2)^2}
 +\frac{55s_1^4s_0^{\ast 3}}{(1-|s_0|^2)^3}\bigg)+\mathcal{O}(\varepsilon)\,,
\nonumber\\
\mathcal{S}_5&=\frac{1}{(1-|s_0|^2)^6}
 \bigg(s_5+\frac{6(2s_4s_1+2s_3s_2)s_0^{\ast}}{1-|s_0|^2}
 +\frac{26(3s_3s_1^2+3s_2^2s_1)s_0^{\ast 2}}{(1-|s_0|^2)^2}
\nonumber\\
&\quad{}
 +\frac{91\cdot4s_2s_1^3s_0^{\ast 3}}{(1-|s_0|^2)^3} +\frac{273s_1^5s_0^{\ast 4}}{(1-|s_0|^2)^4}\bigg)+\mathcal{O}(\varepsilon)\,,
\nonumber\\
&\dots\,,
\nonumber\\
\mathcal{S}_j&=\frac{1}{(1-|s_0|^2)^{j+1}}\bigg(s_j +\frac{\alpha_{j2}s_0^{\ast}}{1-|s_0|^2}\!\sum_{j_1+j_2=j}\!\!s_{j_1}s_{j_2}
 +\frac{\alpha_{j3}s_0^{\ast 2}}{(1-|s_0|^2)^2}\!\sum_{j_1+j_2+j_3=j}\!\!\!s_{j_1}s_{j_2}s_{j_3}
\nonumber\\
&{} +...+\frac{\alpha_{jk}s_0^{\ast k-1}}{(1-|s_0|^2)^{k-1}}\!\!\sum_{j_1+j_2+...+j_k=j}\!\!\!\!s_{j_1}s_{j_2}...\,s_{j_k} +...
 +\frac{\alpha_{jj}s_0^{\ast j-1}}{(1-|s_0|^2)^{j-1}}s_1^j\bigg)+\mathcal{O}(\varepsilon)\,,
\nonumber\\
&\dots\,,
\label{eqr09}
\end{align}
where
$\alpha_{mk}=\frac{(2m+k)!}{k!(2m+1)!}$. The expression for $\mathcal{S}_j$ was validated by calculations for up to $j=10$.

\subsubsection{Inverted dependence: $\{\mathcal{K}_m\}\mapsto\{K_m\}$}
From Eq.~(\ref{eq104}), similarly to Eq.~(\ref{eqr00}),
\[
\int\mathrm{d}\varphi\,w(\varphi)\,e^{im\varphi} =\int\mathrm{d}\psi\,W(\psi) \left(\frac{e^{i\psi}+z}{1+z^\ast e^{i\psi}}\right)^m\,.
\]
This relation differs from the case $\{a_m\}\mapsto\{A_m\}$ by the sign of $z$ and the fact that $\mathcal{S}_0=0$. 
Hence,
\[
a_j=\left.\big(1+\sum_{m=2}^{\infty}A_m\widehat{Q}_m\big) \left(\frac{\mathcal{S}_0+z}{1+z^\ast\mathcal{S}_0}\right)^{\!j}\,\right|_{\mathcal{S}_0=0},
\]
where $\widehat{Q}_m\equiv\frac{1}{m!}\left(\frac{\partial}{\partial\mathcal{S}_0}\right)^m$.

With the latter equation, one can obtain
\[
a_1=z+(1-|z|^2)\sum_{m=2}^\infty A_m(-z^\ast)^{m-1}\,,
\]
and, for $j\ge2$,
\begin{align}
a_j&=z^j+\sum_{m=2}^\infty A_m(z^\ast)^{m-j}
 \sum_{l=1}^j C_l^jC_m^{l-1+m}(-1)^{l+m}(1-|z|^2)^l\,.
\nonumber
\end{align}
Notice, here $z$ and the series $\{\mathcal{K}_j|j=2,3,...\}$ are input parameters, which determine the leading order of $\{a_j\}$, including the order parameter $a_1$.

Further, similarly to the previous subsection, one can calculate
\begin{align}
s_0&=z+\mathcal{O}(\varepsilon)\,,
\nonumber\\
s_1&=(1-|z|^2)^2\mathcal{S}_1+\mathcal{O}(\varepsilon)\,,
\nonumber\\
s_2&=(1-|z|^2)^3(\mathcal{S}_2-3z^\ast\mathcal{S}_1^2)+\mathcal{O}(\varepsilon)\,,
\nonumber\\
s_3&=(1-|z|^2)^4(\mathcal{S}_3-8z^\ast\mathcal{S}_2\mathcal{S}_1+12z^{\ast 2}\mathcal{S}_1^3) +\mathcal{O}(\varepsilon)\,,
\nonumber\\
&\dots\,,
\nonumber\\
s_j&=(1-|z|^2)^{j+1}\Big(\mathcal{S}_j -\alpha_{j2}z^\ast\!\!\sum_{j_1+j_2=j}\!\!\mathcal{S}_{j_1}\mathcal{S}_{j_2}
 +\alpha_{j3}z^{\ast 2}\!\!\!\sum_{j_1+j_2+j_3=j}\!\!\!
 \mathcal{S}_{j_1}\mathcal{S}_{j_2}\mathcal{S}_{j_3} +\dots
\nonumber\\
&{}+\alpha_{jk}(-z^\ast)^{k-1}\!\!\!\!\sum_{j_1+j_2+\dots+j_k=j}\!\!\!\!\mathcal{S}_{j_1}\mathcal{S}_{j_2}\dots\mathcal{S}_{j_k} +\dots
 +\alpha_{jj}(-z^\ast)^{m-1}\mathcal{S}_1^m\Big)+\mathcal{O}(\varepsilon)\,,
\nonumber\\
&\dots\,.
\label{eqr10}
\end{align}
Eqs.~(\ref{eqr09}) and (\ref{eqr10}) with $\varepsilon\to0$ are mutually inverse transformations, and this property holds for arbitrary truncation order $n$ (which was also confirmed by numerical calculations for random sets $\{s_j\}$). Noticeably, Eqs.~(\ref{eqr09}) and (\ref{eqr10}) take much more sophisticated forms in terms of $K_j$ and $\mathcal{K}_j$ (the sums over $j_1,j_2,\dots,j_k$ acquire lengthy coefficients).

\section{Conclusion}
\label{sec:concl}
We have derived the relationships between the distribution of Watanabe--Strogatz variables $\psi_k$ and circular cumulants of phases $\varphi_k$. The WS transformation parameter $z$ and Fourier amplitudes $A_j$ of distribution $W(\psi)$ are determined by Eq.~(\ref{eqr04}) and Eqs.~(\ref{eqr01})--(\ref{eqr03}), respectively, without any assumptions on values of circular cumulants $\varkappa_j$. Further results pertain to an important case of the hierarchy of circular cumulants $\varkappa_j=\varepsilon^{j-1}s_{j-1}$, which corresponds to the evolution of the ensemble within the $\varepsilon$-vicinity of the Ott--Antonsen solution~\cite{Tyulkina-etal-2018}.\footnote{Strictly speaking, one can make two rigorous claims here. (i)~In the case of intrinsic noise of intensity $\sigma^2=\varepsilon$, the perturbed OA solution is a hierarchy $\varkappa_j\propto\varepsilon^{j-1}$. (ii)~If the intrinsic noise intensity does not exceed $\varepsilon$ and the initial state has the form of hierarchy $\varkappa_j\propto\varepsilon^{j-1}$, the hierarchy persists on timescales $\mathcal{O}(1)$.}
For the hierarchy of circular cumulants of $\varphi_k$, which generally does not require hierarchy of amplitudes $a_j=\langle{e^{ij\varphi}}\rangle$ with a small parameter, amplitudes $A_j=\langle{e^{ij\psi}}\rangle$ obey the hierarchy $A_{2m-1},A_{2m}\propto\varepsilon^m$ [see Eq.~(\ref{eqr08})]. However, $\{A_j\}$ is a poor representative of the ensemble state, since the leading order of $A_j$ is determined solely by $\varkappa_1=s_0$ and $\varkappa_2=\varepsilon s_1$ and, moreover, the leading terms of $A_j$ mutually cancel in the expressions for high-order cumulants of $\psi$. The circular cumulants of $\psi$ and $\varphi$ obey the same hierarchy; to the leading order, their mutual transforms are determined by Eqs.~(\ref{eqr09}) and (\ref{eqr10}). These transforms remain exactly mutually inverse for arbitrary order of their truncation.

Employing the reported relations between circular cumulants and the probability density of WS variables $\psi_k$, one can interpret the results obtained with circular cumulants within the framework of the WS approach. For instance, the effect of intrinsic noise on the chimera states in the system of two hierarchically coupled oscillator populations~\cite{Abrams-etal-2008} was studied in~\cite{Tyulkina-etal-2018} in terms of circular cumulants. The noise was found to make the states of the partially synchronous population of the form $\varkappa_j\sim\sigma^{2(j-1)}$ attracting. For the states of such a form, according to the reported relations, the density $W(\psi)=(2\pi)^{-1}+\mathcal{O}(\sigma^2)$. One gains an interpretation that a weak intrinsic noise breaks the nondissipativeness of the dynamics of $W(\psi,t)$ and makes a slightly nonuniform distribution of $\psi_k$ attracting. On the other hand, for specific deterministic violations of the OA form, one may have intuition on the dynamics of the WS variables. With known relations between $W(\psi)$ and circular cumulants, this intuition can guide the search for an approach to dealing with the problem in terms of circular cumulants.

\section*{Acknowledgements}
The author is thankful to Arkady Pikovsky for drawing his attention to the subject of this paper, fruitful discussions and comments.

\appendix
\section{Derivation}
\label{sec:deriv}
\subsection{Calculation of the transform from Kuramoto--Daido order parameters $a_j$ to order parameters $A_j$ of WS variables}
\label{sec:der-Aj}
From~(\ref{eqr00}) with $W(\psi)=\frac{1}{2\pi}[1+\sum_j(A_je^{-ij\psi}+c.c.)]$, one finds expansions in $z$:
\begin{align}
A_1&=-z+(1-|z|^2)(a_1+z^\ast a_2 +z^{\ast 2}a_3 +z^{\ast 3}a_4+z^{\ast 4}a_5+z^{\ast 5}a_6+\dots)
\label{eqab01}
\\
A_2&=z^2+(1-|z|^2)\big(-2za_1 +(1-3|z|^2)a_2 +(2-4|z|^2)z^{\ast}a_3
\nonumber\\
&\qquad
{}+(3-5|z|^2)z^{\ast{2}}a_4 +(4-6|z|^2)z^{\ast{3}}a_5 +(5-7|z|^2)z^{\ast{4}}a_6 +\dots\big)\,,
\label{eqab02}
\\
A_3&=-z^3+(1-|z|^2)\big(3z^2a_1 +(-3+6|z|^2)za_2 +(1-8|z|^2+10|z|^4)a_3
\nonumber\\
&\qquad
{}+(3-15|z|^2+15|z|^4)z^{\ast}a_4 +(6-24|z|^2+21|z|^4)z^{\ast{2}}a_5
\nonumber\\
&\qquad
{}+(10-35|z|^2+28|z|^4)z^{\ast{3}}a_6+\dots\big)\,,
\label{eqab023}
\\
&\dots\,.
\nonumber
\end{align}

For calculations we formally use $\varkappa_j=\varepsilon^{j-1}s_{j-1}$, although the final results of this subsection are present in the form which is free from the formal small parameter $\varepsilon$. With moments $a_j$ expressed via cumulants $\varkappa_j$, Eq.~\eqref{eqab01} can be expanded:
\begin{align}
A_1&\textstyle
=-z+(1-|z|^2)(s_0+z^\ast s_0^2 +z^{\ast 2}s_0^3 +z^{\ast 3}s_0^4+z^{\ast 4}s_0^5+\dots)
\nonumber\\
&\textstyle{}
+\varepsilon z^\ast s_1(1-|z|^2)\big(1+3z^{\ast}s_0 +6(z^{\ast}s_0)^2 
+\dots +\frac{(m+2)!}{2!\,m!}(z^{\ast}s_0)^m+\dots\big)
\nonumber\\
&\textstyle{}
+\varepsilon^2(1-|z|^2)\Big[2z^{\ast 2}s_2\big(1+4z^{\ast}s_0 +10(z^{\ast}s_0)^2 
+\dots +\frac{(m+3)!}{3!\,m!}(z^{\ast}s_0)^m+\dots\big)
\nonumber\\
&\textstyle{}\qquad
+3z^{\ast 3}s_1^2\big(1+5z^{\ast}s_0 +15(z^{\ast}s_0)^2
+\dots +\frac{(m+4)!}{4!\,m!}(z^{\ast}s_0)^m+\dots\big)\Big]
\nonumber\\
&\textstyle{}
+\varepsilon^3(1-|z|^2)\Big[6z^{\ast 3}s_3\big(1+5z^{\ast}s_0 +15(z^{\ast}s_0)^2 
+\dots +\frac{(m+4)!}{m!\,4!}(z^{\ast}s_0)^m+\dots\big)
\nonumber\\
&\textstyle{}\qquad
+20z^{\ast 4}s_2s_1\big(1+6z^{\ast}s_0 +21(z^{\ast}s_0)^2 
+\dots +\frac{(m+5)!}{m!\,5!}(z^{\ast}s_0)^m+\dots\big)
\nonumber\\
&\textstyle{}\qquad
+15z^{\ast 5}s_1^3\big(1+7z^{\ast}s_0 +28(z^{\ast}s_0)^2 +\dots +\frac{(m+6)!}{m!\,6!}(z^{\ast}s_0)^m+\dots\big)\Big]
\nonumber\\
&\textstyle{}
+\varepsilon^4(1-|z|^2)\Big[24z^{\ast 4}s_4\big(1+6z^{\ast}s_0 
+\dots\big)
+90z^{\ast 5}s_3s_1\big(1+7z^{\ast}s_0 
+\dots\big)
\nonumber\\
&\textstyle{}\qquad
+40z^{\ast 5}s_2^2\big(1+7z^{\ast}s_0 
+\dots\big)
+210z^{\ast 6}s_2s_1^2\big(1+8z^{\ast}s_0 
+\dots\big)
\nonumber\\
&\textstyle{}\qquad
+105z^{\ast 7}s_1^4\big(1+9z^{\ast}s_0 
+\dots\big)\Big]+\dots
\nonumber\\
&{}
=-z+(1-|z|^2)\Big[s_0F_0 +\varepsilon z^\ast s_1F_2
 +\varepsilon^2\big(2z^{\ast 2}s_2F_3 +3z^{\ast 3}s_1^2F_4\big)
 +\varepsilon^3\big(6z^{\ast 3}s_3F_4
\nonumber\\
&\qquad{}
 +20z^{\ast 4}s_2s_1F_5 +15z^{\ast 5}s_1^3F_6\big)
 +\varepsilon^4\big(24z^{\ast 4}s_4F_5 +90z^{\ast 5}s_3s_1F_6 +40z^{\ast 5}s_2^2F_6
\nonumber\\
&\qquad{}
 +210z^{\ast 6}s_2s_1^2F_7 +105z^{\ast 7}s_1^4F_8\big)+\dots\Big]\,,
\label{eqab04}
\end{align}
where
$F_m\equiv\left.\frac{1}{m!}\frac{\mathrm{d}^m}{\mathrm{d}\xi^m}\frac{1}{1-\xi}\right|_{\xi=z^{\ast}s_0}
=\frac{1}{(1-z^{\ast}s_0)^{m+1}}$.

One can write
$A_j=A_j^{(0)}+\varepsilon A_j^{(1)}+\varepsilon^2A_j^{(2)}+\dots$
(notice, the expansion in $A_j^{(n)}$ is not a true expansion with respect to $\varepsilon$, because for a proper expansion in $\varepsilon$, $z$ is to be expanded as well). Hence, Eq.~\eqref{eqab04} yields
\begin{align}
A_1^{(0)}&=-z+(1-|z|^2)\frac{s_0}{1-z^\ast s_0}\,,
\label{eqab:a10}\\
A_1^{(1)}&=s_1\widehat{Q}_2A_1^{(0)},
\label{eqab:a11}\\
A_1^{(2)}&=(2s_2\widehat{Q}_3+3s_1^2\widehat{Q}_4)A_1^{(0)},
\label{eqab:a12}\\
A_1^{(3)}&=(6s_3\widehat{Q}_4+20s_2s_1\widehat{Q}_5+15s_1^3\widehat{Q}_6)A_1^{(0)},
\label{eqab:a13}\\
A_1^{(4)}&=(24s_4\widehat{Q}_5+90s_3s_1\widehat{Q}_6+40s_2^2\widehat{Q}_6
 +210s_2s_1^2\widehat{Q}_7+105s_1^4\widehat{Q}_8)A_1^{(0)},
\label{eqab:a14}\\
&\dots\,,
\nonumber
\end{align}
where
$\widehat{Q}_m\equiv\frac{1}{m!}\left(\frac{\partial}{\partial s_0}\right)^m$.

One can notice, that Eqs.~\eqref{eqab:a11}--\eqref{eqab:a14} summed up in $A_1$ form the groups, where one can recognize $p_m$ defined by Eqs.~(\ref{eqr02}). One finds
$A_1=\big(1+\sum_{m=2}^\infty p_m\widehat{Q}_m\big)A_1^{(0)}$, which corresponds to Eqs.~(\ref{eqr01}), (\ref{eqr03}).

From Eq.~\eqref{eqab02}, similarly to the case of $A_1$, one can obtain
\begin{align}
A_2^{(0)}
&=z^2+(2-2|z|^2)zs_0+(1-4|z|^2+3|z|^4)s_0^2 +(2-6|z|^2+4|z|^2)z^{\ast}s_0^3
\nonumber\\
&+(3-8|z|^2+5|z|^4)z^{\ast 2}s_0^4 +(4-10|z|^2+6|z|^4)z^{\ast 3}s_0^5 +\dots
\nonumber\\
&=(s_0-z)^2(1+2z^\ast s_0+3(z^\ast s_0)^2 +4(z^\ast s_0)^3 +5(z^\ast s_0)^4+\dots)
\nonumber\\
&=(s_0-z)^2\left.\frac{\mathrm{d}}{\mathrm{d}\xi}\frac{1}{1-\xi}\right|_{\xi=z^{\ast}s_0}
 =\frac{(s_0-z)^2}{(1-z^{\ast}s_0)^2}\,.
\nonumber
\end{align}
\begin{align}
A_2^{(1)}&=(1-4|z|^2+3|z|^4)s_1 +3(2-6|z|^2+4|z|^2)z^{\ast}s_0s_1
\nonumber\\
&+6(3-8|z|^2+5|z|^4)z^{\ast 2}s_0^2s_1 +10(4-10|z|^2+6|z|^4)z^{\ast 3}s_0^3s_1 +\dots
\nonumber\\
&=\frac{s_1}{2}\frac{\partial^2}{\partial s_0^2}A_2^{(0)}\,.
\nonumber
\end{align}
\begin{align}
A_2^{(2)}&=(2-6|z|^2+4|z|^2)z^{\ast}(2s_2) +(3-8|z|^2+5|z|^4)z^{\ast 2}(8s_0s_2+3s_1^2)
\nonumber\\
&\qquad
{}+(4-10|z|^2+6|z|^4)z^{\ast 3}(20s_0^2s_2+15s_0s_1^2)+\dots
\nonumber\\
&=\frac{2s_2}{3!}\frac{\partial^3}{\partial s_0^3}A_2^{(0)}
+\frac{3s_1^2}{4!}\frac{\partial^4}{\partial s_0^4}A_2^{(0)}\,.
\nonumber
\end{align}
\[
A_2^{(3)}=\dots=\frac{6s_3}{4!}\frac{\partial^4}{\partial s_0^4}A_2^{(0)}
+\frac{20s_2s_1}{5!}\frac{\partial^5}{\partial s_0^5}A_2^{(0)}
+\frac{15s_1^3}{6!}\frac{\partial^6}{\partial s_0^6}A_2^{(0)}\,.
\]

Similarly, for $A_3$,
\begin{align}
A_3^{(0)}
&=(s_0-z)^3\big(1+3\xi+6\xi^2+10\xi^3+15\xi^4+\dots\big)|_{\xi=z^\ast s_0}
\nonumber\\
&=(s_0-z)^3\left.\frac{1}{2}\frac{\mathrm{d}^2}{\mathrm{d}\xi^2}\frac{1}{1-\xi}\right|_{\xi=z^{\ast}s_0}
=\frac{(s_0-z)^3}{(1-z^{\ast}s_0)^3}\,.
\nonumber
\end{align}
\begin{align}
A_3^{(1)}
&=\frac{s_1}{2}\frac{\partial^2}{\partial s_0^2}A_3^{(0)}\,,
\qquad
A_3^{(2)}
=\frac{2s_2}{3!}\frac{\partial^3}{\partial s_0^3}A_3^{(0)}
+\frac{3s_1^2}{4!}\frac{\partial^4}{\partial s_0^4}A_3^{(0)}\,,
\nonumber
\end{align}
\[
A_3^{(3)}
=\frac{6s_3}{4!}\frac{\partial^4}{\partial s_0^4}A_3^{(0)}
+\frac{20s_2s_1}{5!}\frac{\partial^5}{\partial s_0^5}A_3^{(0)}
+\frac{15s_1^3}{6!}\frac{\partial^6}{\partial s_0^6}A_3^{(0)}\,.
\]

The results for $A_1$, $A_2$, $A_3$ and similar calculations for $A_4$ and $A_5$ can be written in a general form of Eqs.~(\ref{eqr01}) and (\ref{eqr03}). Thus, we conclude that Eqs.~(\ref{eqr01}) and (\ref{eqr03}) are valid for all $A_j$.

\subsection{Calculation of $\widehat{Q}_{j}A_j^{(0)}$ and $\widehat{Q}_{j+1}A_j^{(0)}$}
\label{sec:der-QmAj}
For $m=j$, with recurrence relation \eqref{eqr06}, there is only one possible route from $A_j^{(0)}$ to $A_0^{(0)}$ in $\widehat{Q}_mA_j^{(0)}$: all the moves are $j'\to j'-1$, yielding multipliers $\frac{j'}{1-|z|^2}$. Hence,
\[
\frac{1}{j!}\frac{\partial^j}{\partial s_0^j}A_j^{(0)} =\frac{1}{j!}\frac{j!}{(1-|z|^2)^j}A_0^{(0)} =\frac{1}{(1-|z|^2)^j}\,.
\]
For $m=j+1$, on the route from $A_j^{(0)}$ to $A_0^{(0)}$ there must be one move $j'\to j'$, which yields multiplier $2j'z^\ast\frac{1}{1-|z|^2}$, while all the moves $j''\to j''-1$ yield multipliers $\frac{j''}{1-|z|^2}$, which combine into $\frac{j!}{(1-|z|^2)^j}$ for $j''=j,j-1,...,1$. One has to sum over possible sites of the moves $j'\to j'$;
\begin{align}
\frac{1}{(j+1)!}\frac{\partial^{j+1}}{\partial s_0^{j+1}}A_j^{(0)}&=\frac{1}{(j+1)!}\sum_{j'=1}^jj'\frac{2z^\ast j!}{(1-|z|^2)^{j+1}}A_0^{(0)}
=\frac{j\,z^\ast}{(1-|z|^2)^{j+1}}\,.
\nonumber
\end{align}
For $m=j+2$, the routes with either one move $j'\to j'+1$ or two moves $j'\to j'$ are to be counted. With $j'\to j'+1$, which yields multiplier $\frac{j'z^{\ast 2}}{1-|z|^2}$, $(j+1)$ moves $j''\to j''-1$, which yield combined multiplier $\frac{j\cdot(j-1)\cdot...\cdot (j'+1)}{(1-|z|^2)^{j-j'}}\frac{(j'+1)!}{(1-|z|^2)^{j'+1}}$ on two patches of the route, before and after $j'\to j'+1$, yield the contribution $\sum_{j'=1}^{j}\frac{j'z^{\ast 2}(j'+1)j!}{(j+2)!(1-|z|^2)^{j+2}}A_0^{(0)}=\frac{j\,z^{\ast 2}}{3(1-|z|^2)^{j+2}}$. With two moves $j'_1\to j'_1$ and $j'_2\to j'_2$, which yield multipliers $\frac{2j'_1z^\ast}{1-|z|^2}$ and $\frac{2j'_2z^\ast}{1-|z|^2}$, and $j$ moves $j''\to j''-1$, which yield combined multiplier $\frac{j!}{(1-|z|^2)^j}$, one obtains the contribution $\sum_{j'_1=1}^j\sum_{j'_2=1}^j\frac{4j'_1j'_2z^{\ast 2}j!}{(j+2)!(1-|z|^2)^{j+2}}A_0^{(0)} =\left(\sum_{j'_1=1}^jj'_1\right)^2\frac{4z^{\ast 2}}{(j+1)(j+2)(1-|z|^2)^{j+2}}$. Summing up,
\[
\frac{1}{(j+2)!}\frac{\partial^{j+2}}{\partial s_0^{j+2}}A_j^{(0)}
 =\frac{j(j^2+\frac43 j+\frac23)z^{\ast 2}}{(j+2)(1-|z|^2)^{j+2}}.
\]

\end{document}